\pdfoutput=1
\documentclass[reprint,linenumbers,aip, 
 amsmath]{revtex4-1}
\usepackage{graphicx}
\usepackage{dcolumn}
\usepackage{bm}
\usepackage{bm}
\usepackage{hyperref}
\hypersetup{
   colorlinks,=true,
    citecolor=blue,
    filecolor=blue,
    linkcolor= blue,
   urlcolor=blue
}

\begin{document}

\title{Interband absorption in single layer hexagonal boron nitride } 


\author{Aniruddha Konar}
\email[]{akonar@nd.edu}
\affiliation{ Department of Physics, University of Notre Dame, Notre Dame, USA 46556}
\affiliation{ Midwest Institue of Nanoelectronics Discovery (MIND), Notre Dame, USA 46556}
\author{Raj Jana}
\affiliation{  Department of Electrical Engineering, University of Notre Dame, Notre Dame, USA 46556}
\author{Tian Fang}
\affiliation{  Department of Electrical Engineering, University of Notre Dame, Notre Dame, USA 46556}
\affiliation{ Midwest Institue of Nanoelectronics Discovery (MIND), Notre Dame, USA 46556}
\author{Guowang Li}
\affiliation{  Department of Electrical Engineering, University of Notre Dame, Notre Dame, USA 46556}
\author{William O'Brien}
\affiliation{  Department of Electrical Engineering, University of Notre Dame, Notre Dame, USA 46556}
\author{ Debdeep Jena}
\affiliation{ Department of Electrical Engineering, University of Notre Dame, Notre Dame, USA 46556}
\affiliation{ Midwest Institue of Nanoelectronics Discovery (MIND), Notre Dame, USA 46556}

\date{\today}

\begin{abstract}
Monolayer of hexagonal boron nitride (h-BN), commonly known as ``white graphene'' is a promising wide bandgap semiconducting material for deep-ultaviolet optoelectronic devices. In this report, the light absorption of a single layer hexagonal boron nitride is calculated using a tight-binding Hamiltonian. The absorption is found to be monotonically decreasing function of photon energy compared to graphene  where absorption coefficient is independent of photon energy and  characterized by the effective fine-structure constant.
\end{abstract}

\maketitle 
Wide bandgap ($E_{g}$) III-nitride materials and related compounds are being investigated extensively for high speed optoelectronic devices in the visible and ultraviolet (UV) range of the electromagnetic spectrum\cite{WaltereitNature00,NakamuraJJAPL96}. For example, gallium nitride (GaN) and related compounds have been  used to fabricate high-speed blue-ray laser based devices\cite{NakamuraJJAPL96,NakamuraAPL98}. Moreover, the increasing demand of shorter wavelength-based (in UV range) devices for optical storage, environmental protection and medical treatment has pushed the researcher look for materials with bandgap higher than of GaN. One  such wide-bandgap material is hexagonal boron nitride\cite{WantanabeNature04,KubotaScience04} with $E_{g}\sim5.8$ eV. \\
\par
 Hexagonal boron nitride is a layered material where each layer is comprised off honeycomb arrangement of alternation boron and nitrogen atoms covalently bonded to each other via sp$^{2}$ hybridization whereas, the layers are held together by weak Van del Wall force. Few layers of single crystal h-BN has already been achieved by mechanical cleavage\cite{PacileAPL08,DeanNature10} as well as by chemical vapor deposition\cite{SongNL2010}. Moreover, these few layers can be thinned down to atomically smooth flat single layer h-BN by selective chemical reaction\cite{AuwarterChem04,NagNano10}  or by careful micro-mechanical cleavage\cite{DeanNature10,Gorbachevarxive}. Unlike graphene ($E_{g}=0$), the onsite potential difference between the boron and nitrogen atoms in the unit cell gives rise large bandgap in monolayer h-BN making it a potential two dimensional (2D) semiconducting material for UV-based optical devices. In this work, starting with a simple tight-binding Hamiltonian, we derive an effective low-energy Hamiltonian near the band edge at certain symmetry points of the Brillouin zone. Using this effective Hamiltonian, we derive the carrier-photon interaction in a monolayer h-BN and calculate the interband absorption of photon - an important parameter for optoelectronic devices (laser, photodiode and phototransistors etc). 

\par
\begin{figure}[t]%
\includegraphics*[width=80 mm]{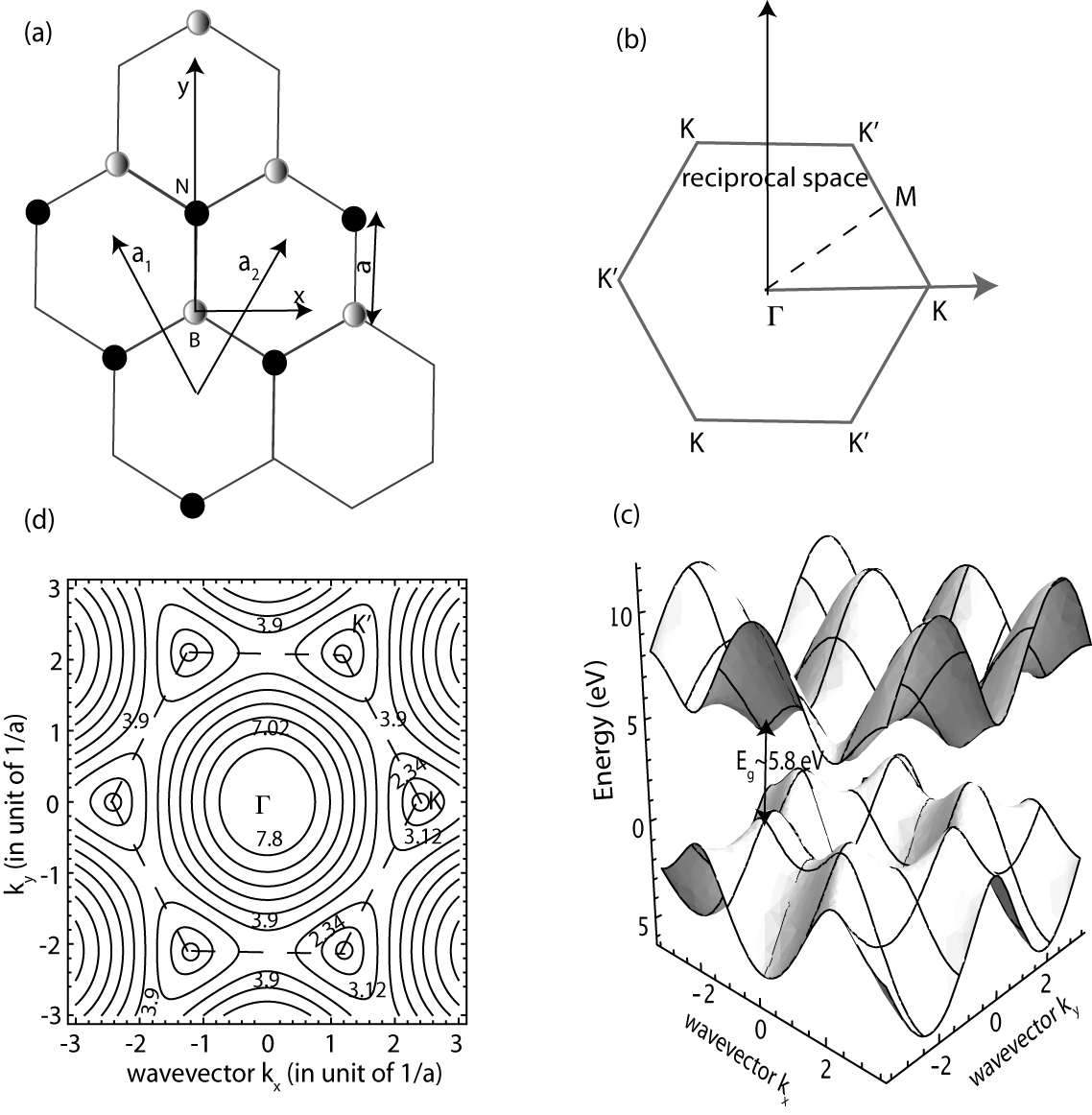}
\caption{%
  a) 2D honecomb lattice of h-BN , b) constant energy contours of h-BN. The numbers on the contour lines corresponds to the energies compare to conduction band edge at ${\mathcal{K}}$ and ${\mathcal{K'}}$ points. }
\label{Fig1}
\end{figure}
 We consider an infinite sheet of monolayer h-BN of lattice constant $a$ as shown in Fig.\ref{Fig1}a). Denoting the primitive vectors ${\bf{a_{1}}}=( \sqrt{3}/2, 3/2)$ and ${\bf{a_{1}}}=(- \sqrt{3}/2, 3/2)$ and accounting the nearest neighbor interaction (minimal tight binding model), the tight-binding Hamiltonian of h-BN sheet is given by
 \begin{equation}
 {\mathcal{H}}=  \left(\begin{array}{cc}
    \epsilon_{B} & h(k) \\ 
    h(k)^{\star} & \epsilon_{N} \\ 
  \end{array}\right)
 \end{equation}
 where, $\epsilon_{B} (\epsilon_{N})$ are the onsite energy of boron (nitrogen) atom and $h(k)=-t\left[1+e^{i{\bf{k}}\cdot{\bf{a_{1}}}}+e^{i{\bf{k}}\cdot{\bf{a_{2}}}}\right]$, $t$ being the hoping energy between neighboring boron and nitrogen atoms. The energy eigenvalues corresponding to the above Hamiltonian is 
 \begin{widetext}
 \begin{equation}
 {\mathcal{E}}_{\pm}(k_{x},k_{y})=A\pm\sqrt{B^{2}+t^{2}\left[1+4\cos (3k_{x}a/2)\cos (\sqrt{3}k_{y}a/2)+4\cos^{2}(\sqrt{3}k_{y}a/2)\right]},
 \end{equation}
 \end{widetext}
  where $A=(\epsilon_{B}+\epsilon_{N})/2$, 
 $B=(\epsilon_{B}-\epsilon_{N})/2$ and $\pm$ corresponds to the upper (conduction) and lower (valence) branch of the bandstructure. The energy eigenvalues differ from graphene by the nonzero onsite potential difference between boron and nitrogen atoms characterized by the parameter $B$. Figure \ref{Fig1}d) shows the constant energy lines of the conduction band in the momentum space. The Brillouin zone contains two equivalent valleys (${\mathcal{K}}$ and ${\mathcal{K'}}$) where band extrema occurs with a bandgap $E_{g}={\mathcal{E}}_{+}-{\mathcal{E}}_{-}=2B=5.8$ eV as shown in Fig.\ref{Fig1}c). For low values of $k$ ($|ka|\ll 1$) around ${\mathcal{K}}$ (${\mathcal{K'}}$ shown in Fig.\ref{Fig1}b) point, the energy dispersion relation to the lowest order of $k$ can be approximated as
 \begin{eqnarray}
 {\mathcal{E}}_{c}(k)&=& E_{g}+\frac{\hbar^{2}k^{2}}{2m^{\star}} ,  \ \ \ \ \ \ \ \ \ \ \mbox{(conduction band)} \nonumber \\
{\mathcal{E}}_{v}(k)&=&  - \frac{\hbar^{2}k^{2}}{2m^{\star}}, \ \ \ \ \mbox{(valence band)}
\end{eqnarray}
Where, $\hbar$ is the reduced Planck constant and $m^{\star}=2\hbar^{2}E_{g}/(9a^{2}t^{2})$ is the effective mass of carriers. For $E_{g}=5.8$ eV, $t=2.92$ eV\cite{RobertsonPRB84} and $a\sim 1.5\AA\ $, the effective mass for carriers (both electron and hole) is $m^{\star}\sim 0.6 m_{0}$, where $m_{0}$ is the bare electron mass. It should be noted that this parabolic isotropic energy dispersion (as shown in Fig.\ref{Fig1}e) is valid for small energies (up to 200 meV) near the band edges. At high energies, parabolic approximation fails and bandstructure is highly anisotropic (trigonal warping) as shown by the energy contours in Fig. \ref{Fig1}d). The effective low-energy Hamiltonian around ${\mathcal{K}}$ point can be obtained by expanding the original Hamiltonian for small $k$ and to the lowest order of $k$ is given by
 \begin{eqnarray}
 {\mathcal{H}}_{eff}&=&  \left(\begin{array}{cc}
    \epsilon_{B} & {\mathcal{\pi}} \\ 
    {\mathcal{\pi}}^{\dagger} & \epsilon_{N} \\ 
  \end{array}\right)\nonumber\\
 &=& \left(\begin{array}{cc}
    \epsilon_{B} & 0 \\ 
    0 & \epsilon_{N} \\ 
  \end{array}\right)+v_{F}\left[{\boldsymbol{{\sigma\cdot p}}}\right]
 \end{eqnarray}
where, $v_{F}= (3at)/2\hbar$, a characteristic velocity in analogy to Fermi velocity in graphene, ${\mathcal{\pi}}=\hbar v_{F}(k_{x}+ik_{y})$, {\boldmath{$\sigma$}}=$(\sigma_{x},\sigma_{y})$ is the Pauli spin matrices\cite{GriffithsQM}, {\boldmath{$p$}}=$\hbar${\bf{k}} is the momentum. Note that for same atoms in the unit cell ($\epsilon_{B}=\epsilon_{N}$), the effective Hamiltonian maps to the graphene Hamiltonian. Neglecting terms in $k^{2}$ and higher orders,  corresponding eigenvectors at ${\mathcal{K}}$ point can be written as
\begin{equation}
{\mathcal{V}}_{C}^{{\mathcal{K}}}(k)=\left(\begin{array}{cc}
e^{i\theta_{k}}\\ \\
\frac{\hbar v_{F}|k|}{E_{g}}\\
\end{array}\right), \ \ \mbox{and} \ \
{\mathcal{V}}_{V}^{{\mathcal{K}}}(k)=\left(\begin{array}{cc}
\frac{-\hbar v_{F}|k|}{E_{g}}\\ \\
e^{-i\theta_{k}}\\
\end{array}\right)
\end{equation}
where, C( V) stands for conduction (valence) band and $\theta_{k}=\tan^{-1}(k_{y}/k_{x})$. Note that  $\langle {\mathcal{V}}_{C}^{{\mathcal{K}}}|{\mathcal{V}}_{V}^{{\mathcal{K}}}\rangle=0$ and $\langle{\mathcal{V}}_{C}^{{\mathcal{K}}}|{\mathcal{V}}_{C}^{{\mathcal{K}}}\rangle\approx 1$ as $(\hbar v_{F}k/E_{g})^{2}\approx 0$ in the long wavelength limit. The corresponding wavefunction is then given by $\Psi_{C/V}(k)=\left[{\mathcal{U}}(r){\mathcal{V}}_{C/V}^{{\mathcal{K}}}e^{i{\bf{k}}\cdot{\bf{r}}}\right]/\sqrt{S}$, where ${\mathcal{U}}(r)$ is the Bloch function given by linear combination of $p_{z}$ orbitals of boron and nitrogen atoms in the unit cell.  A similar Hamiltonian and eigenvectors can be derived at other inequivalent valley ${\mathcal{K}}'$.\\
\par
\begin{figure}[t]%
\includegraphics*[width=88 mm]{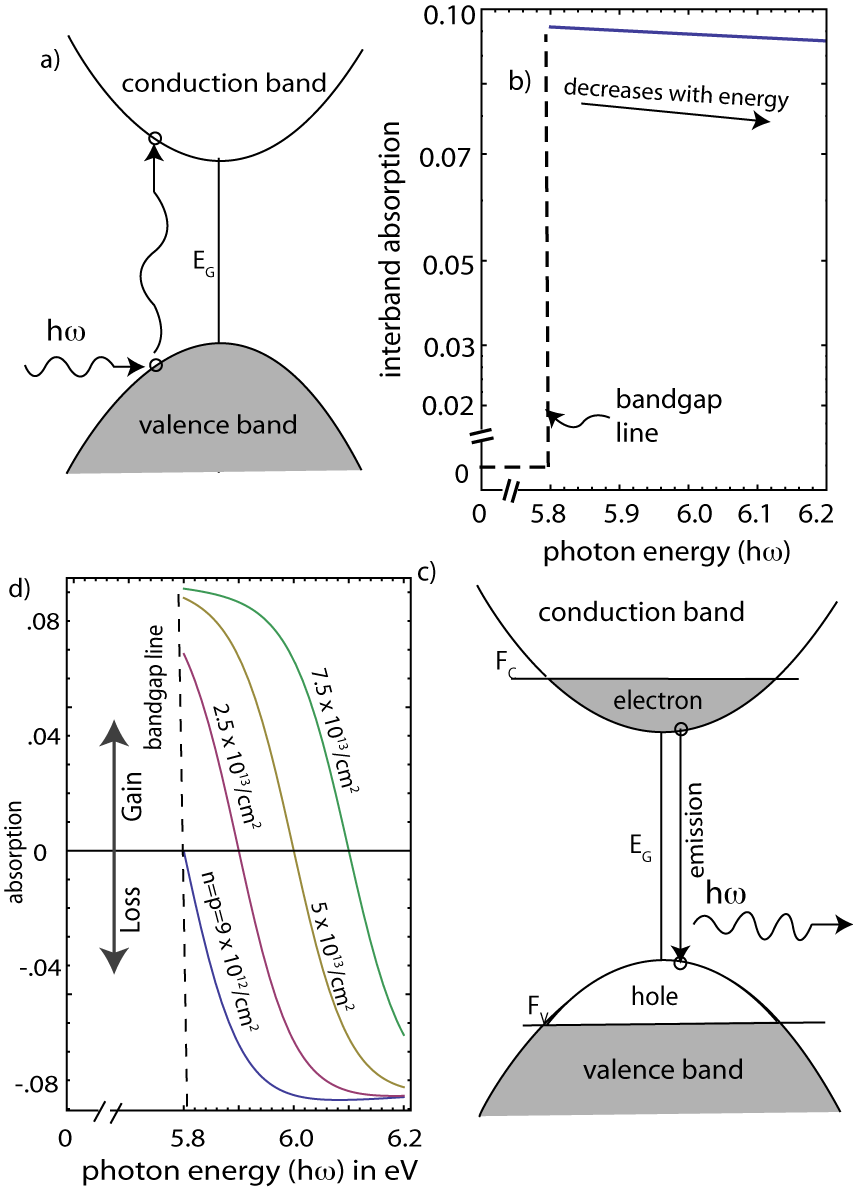}
\caption{%
  a) schematic diagram for photon absorption by free electrons in hexagonal boron nitride, b) equilibrium absorption of single layer h-BN as a function of photon energy near the band edges, c)\& d) absorption in non-equilibrium case of population inversion.}
\label{Fig2}
\end{figure}

In the last section we have derived the low-energy Hamiltonian and the two-component wave-vectors of conduction and valence band at ${\mathcal{K}}$ point. Now we are at a position to to derive optical absorption by single layer h-BN for phonon energy close to the band-gap $E_{g}$. We assume a circularly polarized light of electric field ${\bf{E}}=E_{0}(\hat{x}\pm i\hat{y})e^{i{\bf{q}}\cdot{\bf{r}}-\omega t}/\sqrt{2}$ with frequency $\omega$ falls perpendicularly per unit area of h-BN. Here $\pm$ denotes the left or right handed circular polarization and ${\bf{r}}$ is the position vector.  From classical electrodynamics\cite{BookJackson}, the canonical momentum of an electron in an electromagnetic (EM)  field is given by $({\bf{p}}+e{\bf{A}})$, where ${\bf{A}}$ is the vector potential corresponding to the EM field.  Replacing ${\bf{A}}$ by $({\bf{p}}-e{\bf{A}})$ in Eq. 4, the total Hamiltonian in the presence of EM field can be written as ${\mathcal{H}}={\mathcal{H}}_{eff}+{\mathcal{H}}_{int}$, where the interaction Hamiltonian between the carriers with EM field is given by
\begin{equation}
{\mathcal{H}}_{int}=-ev_{F}\left[{\boldsymbol{{\sigma\cdot A }}}\right].
\end{equation}
Note that, for the low-energy Hamiltonian, the coupling of EM field with the carriers in h-BN is similar to graphene\cite{NairScience}. This interaction Hamiltonian acts as a perturbation to the original Hamiltonian and as a result electron makes transitions from valence band to conduction band (see Fig.\ref{Fig2}a) upon absorbing light. If ${\mathcal{M}}(k,k')=\langle\Psi_{C}(k',r)|{\mathcal{H}}_{int}|\Psi_{V}(k,r)\rangle$ is the matrix element of scattering from a state $|k,r\rangle$ in the valence band to a state $|k',r\rangle$ in conduction band, the number of such transition per unit time per unit area is given by Fermi golden rule\cite{GriffithsQM}
\begin{eqnarray}
\eta(\omega)=\frac{2\pi}{\hbar}g_{s}g_{v}\int\int\frac{{\bf{d^{2}k}}}{\left(2\pi\right)^{2}}\frac{{\bf{d^{2}k'}}}{\left(2\pi\right)^{2}}&&\left|{\mathcal{M}}(k,k')\right|^{2}\delta({\mathcal{E}}_{k'}-{\mathcal{E}}_{k}-\hbar\omega)\nonumber \\
&&\times \left[f^{0}_{ev}({\mathcal{E}}_{k}) - f^{0}_{ec}({\mathcal{E}}_{k'})\right]\nonumber\\
=\left(\frac{e^{2}}{\hbar}\right)\frac{E_{g}}{\left(\hbar\omega\right)^2}E_{0}^{2}\Theta\left(\hbar\omega -E_{g}\right),
\end{eqnarray}
where $g_{s}=2$ is the spin degeneracy, $g_{v}=2$ is the valley degeneracy, $\delta(..)$ is the Dirac delta function, $\Theta(x)$ is the unit step function and $f_{ec}^{0}(f_{ev}^{0})$ is the equilibrium occupation probability of electron in conduction (valence) band. In deriving the rate in Eq.7, we assumed a completely filled valence band ($f_{ev}^{0}$=1) and an empty conduction band ($f_{ev}^{0}=0)$ and the form of the vector potential ${\bf{A}}=-i{\bf{E}}/\omega$ is used. The minus sign in the rate signify the energy loss due to absorption. Assuming a single photon of energy $\hbar\omega$ is absorbed per transition, then absorbed energy per unit time per unit area is ${\mathcal{W}}_{a}=\eta(\omega)\hbar\omega$. The incident energy flux is given by Poynting vector ${\mathcal{W}}_{i}=c\epsilon_{0}\left|{\bf{E}}\right|^{2}$, where $c$ is the speed of light in vacuum and $\epsilon_{0}$ is the free-space permittivity.  Then the absorption is given by
\begin{eqnarray}
{\mathcal{P}}(\hbar\omega)&=&\frac{{\mathcal{W}}_{a}}{{\mathcal{W}}_{i}}\nonumber\\
&=&4\pi\alpha\left(\frac{E_{g}}{\hbar\omega}\right)\Theta\left(\hbar\omega -E_{g}\right),
\end{eqnarray}
where $\alpha=e^{2}/(4\pi\epsilon_{0}\hbar c)$ is the fine-structure constant.\\
\par
Fig.\ref{Fig2}b) shows the interband absorption by free carriers of a single layer h-BN as a function of photon energy in equilibrium (i.e. $f_{ev}^{0}$=1 and $f_{cv}^{0}$=0). The absorption is maximum at band edges for $\hbar\omega=E_{g}$ and decreases inversely with photon energies. The maximum absorption is given by ${\mathcal{P}}_{max}=4\pi\alpha = 0.09 $.  This is in sharp contrast with single layer graphene, where absorption is independent of photon energy and given by a constant value ${\mathcal{P}}_{gr}=0.02$ at all photon energies \cite{NairScience}.  Moreover at band edge i.e. for $\hbar\omega=E_{g}$, the absorption of single layer h-BN is independent of material parameters and proportional to the universal fine structure constant.  \\
\par
Under current injection or optical pumping, the system is driven out of equilibrium and  there will be both electrons and holes h-BN layer. In such non-equilibrium situation, the carrier occupation is still given by Fermi function but with two different quasi-Fermi level $F_{c}$ and $F_{v}$\cite{BookChuang} for electrons and holes in conduction and  valence band respectively as shown in Fig.\ref{Fig2}c). Consequently, the absorption coefficient is modified to 
\begin{eqnarray}
{\mathcal{P}}_{ne}(\hbar\omega)&=&{\mathcal{P}}(\hbar\omega)\left[f_{ev}(k_{0})-f_{ec}(k_{0})\right]\nonumber\\ \\
 k_{0}&=&\sqrt{\frac{m^{\star}}{\hbar^{2}}(\hbar\omega-E_{g})},
\end{eqnarray}
where $f_{ec}(f_{ev})$ is the quasi-equilibrium occupation probability of electrons (holes).  Under population inversion ($f_{ec}\gg f_{ev}$), absorption becomes negative and consequently we have gain in the medium. The condition of achieving population inversion is given by well-known Bernard-Duraffourg\cite{BernardPSS61} inequality, i.e. $E_{g}<\hbar\omega<F_{c}-F_{v}$.  Figure 2d) shows absorption of single layer h-BN for two different carrier injections. \\
\par
In conclusion, we have developed carrier-photon interaction in single layer hexagonal boron nitride using a minimal tight binding Hamiltonian for small energies near the band edges. It is found that light  couples linearly with momentum of carrier in boron nitride through the off-diagonal matrix element in the Hamiltonian. For a ``zero'' band gap $sp^{2}$ bonded 2D crystal (graphene) the interband absorption coefficient is given by $\pi\alpha$.  It is shown that the existence of band-gap, interband absorption gets renormalized by the ratio of the bandgap and the incident photon energy ($E_{g}/\hbar\omega$). The interband optical absorption is maximum for the photon energy equal to the band gap and decreases with higher photon energies. For normal incidence of light, single layer h-BN absorbs maximum $9\%$ of light compare to graphene where only $2\%$ \cite{NairScience} of the incident light is absorbed. \\
\par
 The authors would like to acknowledge  National Science Foundation (NSF) NSF Grant Nos. DMR-0907583 and NSF DMR-0645698), Midwest Institute for Nanoelectronics Discovery (MIND) for the financial support for this work.

\end{document}